\documentclass[aps,prd,preprint,nofootinbib]{revtex4}
\usepackage{bbm}
\usepackage{amsfonts}
\usepackage{mathrsfs}
\usepackage{graphicx}
\usepackage{amsmath}
\usepackage{amssymb}
\usepackage{bm}
\usepackage{bbm}
\usepackage{multirow}
\usepackage{epsfig}
\usepackage{graphicx}
\newcommand{\bqa}{\begin{eqnarray}}
\newcommand{\eqa}{\end{eqnarray}}
\newcommand{\beq}{\begin{equation}}
\newcommand{\eeq}{\end{equation}}
\newcommand{ \slashchar }[1]{\setbox0=\hbox{$#1$}   
   \dimen0=\wd0                                     
   \setbox1=\hbox{/} \dimen1=\wd1                   
   \ifdim\dimen0>\dimen1                            
      \rlap{\hbox to \dimen0{\hfil/\hfil}}          
      #1                                            
   \else                                            
      \rlap{\hbox to \dimen1{\hfil$#1$\hfil}}       
      /                                             
   \fi}                                             %
\begin{document}
\title{P-wave Quarkonium Decays to Meson Pairs \\[7mm]}
\author{Long-Bin Chen\footnote{E-mail:
chenglogbin10@mails.gucas.ac.cn} and Cong-Feng Qiao\footnote{E-mail:
qiaocf@gucas.ac.cn }} \affiliation{Department of Physics,
Graduate University of Chinese Academy of Sciences \\
YuQuan Road 19A, Beijing 100049, China}

\author{~\vspace{1.0cm}}


\begin{abstract}
{~\\[-3mm]}The processes of p-wave Quarkonium exclusive decays to two
mesons are investigated, in which the final state vector mesons with
various polarizations are considered separately. In the calculation,
the initial heavy quarkonia are treated in the framework of
non-relativistic quantum chromodynamics, whereas for light mesons,
the light cone distribution amplitudes up to twist-3 are employed.
It turns out that the higher twist contribution is significant and
provides a possible explanation for the observation of the hadron
helicity selection rule violated processes $\chi_{c1}\rightarrow
\phi\phi,\omega\omega$ by the BESIII collaboration in recently. We
also evaluate the $\chi_{b1}\to J/\psi J/\psi$ process and find that
its branching ratio is big enough to be measured at the B-factories.

\vspace {7mm} \noindent {PACS number(s): 12.38.Bx, 13.25.Gv,
14.40.Be }

\end{abstract}
\maketitle

\section{Introduction}
As one of the most interesting fields of high energy physics, the
study of heavy quarkonium plays an important role in understanding
the configuration of hadrons and the non-perturbative behavior of
QCD. Within the quarkonium physics, recently the exclusive double
charmonium production gains much attention in both theory and
experiment. The double charmonium production at the LHC experiment
\cite{lhc} provides an opportunity to test the prevailing effective
theory of quarkonium physics, the non-relativistic
chromodynamics(NRQCD) \cite{NRQCD}, in high energy, and the double
charmonium production at B-factories \cite{dcbf} poses a challenge
to the leading order(LO) perturbative QCD(pQCD) calculation
\cite{Braaten:2002fi,Liu:2002wq,Hagiwara:2003cw}, which stimulated a
series of followup theoretical studies
\cite{Zhang:2005cha,Gong:2007db,Ma:2004qf,Bondar:2004sv,Braguta:2008tg}.

For the double quarkonium production, since generally the
color-octet mechanism \cite{NRQCD} deduced from the NRQCD formalism
is not that crucial as in some other situations, for instance in the
inclusive quarkonium hadroproduction, it is tempting to apply the
light cone formalism to the study \cite{Ma:2004qf}. At the
B-factory, although the center-of-mass energy is only about 10 GeV,
not very high, it is still much larger than the charm quark mass.
And hence, the charm quarks in the double charmonium production may
move on the light cone before hadronization. This fact inspires one
to investigate the double charmonium production processes at the
B-factory energy in the light cone formalism, though the form of
charmonium light cone wave functions are far from mature.

At the B-factory, there are a huge number of bottomnium states, and
the bottomonium decays to charmonium pairs are now under
investigation in experiment \cite{belle}. Theoretically, several
calculations on this issue are performed in the framework of NRQCD
\cite{zdf,yjia,wang,sang}, but only a few are carried on in light
cone formalism. In Refs. \cite{bra1,bra2,shq} the light cone
formalism was applied to calculate the process of bottomonium decays
to double charmonium. The authors of \cite{bra1,shq} found that
higher twist terms were important and the results of
$Br(\eta_b\rightarrow J/\psi+J/\psi)$ are much larger than the
leading order NRQCD \cite{yjia} prediction, which asks for further
studies on quarkonium production and decays in light cone framework.

In the literature, the p-wave charmonia, the $\chi_{cJ}$, exclusive
decays to light meson pairs have been investigated extensively, by
virtue of NRQCD \cite{zdf}, the light cone formalism with
perturbative QCD(pQCD) factorization approach \cite{zhou}, and also
the meson loop technique \cite{qzhao}, respectively. However, the
corresponding study in bottomonium sector, i.e., the study on the
p-wave bottomonia, the $\chi_{bJ}$, exclusive decays to light meson
pairs and double charmonia are very limited. In Refs.
\cite{zdf,bra2}, p-wave bottomonia decays to double $J/\psi$ are
evaluated in light cone formalism at leading order in twist
expansion and in the framework of NRQCD, but the results from
different approaches do not agree with each other.

In this work, we study various processes of p-wave heavy quarkonium
exclusive decays to meson pairs (VV or PP), in which the light cone
distribution amplitudes up to twist-3 are applied to the final
states. Notice that neither relativistic correction in NRQCD, nor
the twist-2 contribution in light cone formalism can explain well
the BESIII experimental data of $\chi_{c1}\rightarrow \phi\phi,\
\omega\omega$ processes, which implies that the helicity selection
rule is seriously violated. We tend to think that the twist-3
contribution is not merely a higher order correction, but rather
significant in the description of these processes.

The rest of this paper is organized as follows. In Section II we
present the strategy and formalism of the study; in Section III, we
perform the numerical calculation and give some discussion on the
results; Section IV is devoted to a summary and conclusions. For the
sake of the reader¡¯s convenience, some of the formulas used are
given in the Appendix.

\section{Calculation Scheme Description and Formalism}\label{II}

\begin{figure}[h]
\begin{center}
\includegraphics[scale=0.4]{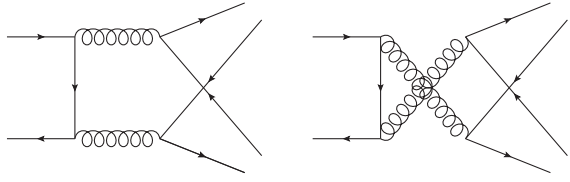}
\caption{ Leading-order QCD Feynman diagrams for $\chi_{QJ} \to$ VV\
or\ PP\ process.  \label{fig1}}
\end{center}
\end{figure}

\subsection{The initial heavy quarkonium}

To begin the calculation, we first determine the character of
initial heavy quarkonium via partons scattering process $Q\bar{Q}
\rightarrow q\bar{q}\ +\ q \bar{q}$, where $Q$ stands for the heavy
quark $b$ or $c$, and $q$ for the constituent quarks of final state
mesons. The schematic Feynman diagrams at the leading order in
perturbative QCD are shown in Figure \ref{fig1}. For initial heavy
quarks $Q$ and $\bar{Q}$, their momenta are assigned as
\bqa p = \frac{P}{2} + l, \ \ \ \ \ \bar{p} = \frac{P}{2} - l, \eqa
respectively. Here, the relative momentum $l$ between heavy quarks
in the center-of-mass system(CMS) is much smaller than the total
momentum $P$ of the quark-antiquark system in the laboratory
system(LS). To construct the spin-triplet states $\chi_{QJ}$, one
needs to project the colors and spinors of $Q$ and $\bar{Q}$ to the
proper quantum number of the states through projection operator by
the standard approach in NRQCD \cite{Braaten:2002fi,Bodwin:2007ga},
that is
\bqa
\Pi_3^\mu\epsilon_\mu&=&-\frac{(\not\!p+m_Q) (\not\!P + 2E_l)
\gamma^\mu(\not\!\bar{p}-m_Q)} {4\sqrt{2}E_l (E_l +
m_Q)}\epsilon_\mu \otimes \frac{\mathbf{1}_c}{\sqrt{N_c}}\ ,
\label{sprojector} \eqa
where $E^2_l = P^2/4 = m_Q^2 - l^2$, $N_c=3$, and $\textbf{1}_c$ is
the unit color matrix. The spin polarization vector $\epsilon_\mu$
satisfies $P\cdot\epsilon=0$ and $\epsilon\cdot\epsilon^*=-1$.

Generally, in the spin-triplet case the expansion of the matrix
element in powers of $l$ has the form \cite{Braaten:2002fi}
\begin{eqnarray}
\mathcal{M}[Q \bar Q(S=1)]&=& \left(
 \mathcal{A}_\rho
+\mathcal{B}_{\rho \sigma} l^\sigma +\mathcal{C}_{\rho \sigma \tau}
l^{\sigma} l^{\tau} + \ldots \right) \epsilon_S^\rho\ .
\label{q-expand:1}
\end{eqnarray}
Then the matrix element for the spin-triplet P-wave heavy quarkonium
states can be expressed as:
\begin{subequations}
\begin{eqnarray}
\mathcal{M}[\chi_{Q0}] &=& \left(\frac{ \langle O_1
\rangle_{\chi_{Q0}} }{ 2 N_c m_Q } \right)^{1/2}
\mathcal{B}_{\rho\sigma} \; \frac{1}{\sqrt{3}} \, I^{\rho\sigma}\ ,
\label{M-3P0}
\\
\mathcal{M}[\chi_{Q1}(\lambda)] &=& \left(\frac{ \langle O_1
\rangle_{\chi_{Q1}} }{ 2 N_c m_Q } \right)^{1/2}
\mathcal{B}_{\rho\sigma} \; \frac{i}{2m_Q\sqrt{2}} \,
\epsilon^{\rho\sigma \lambda \mu}P_{\lambda}
    \epsilon_\mu(\lambda)\ ,
\label{M-3P1}
\\
\mathcal{M}[\chi_{Q2}(\lambda)] &=& \left(\frac{ \langle O_1
\rangle_{\chi_{Q2}} }{ 2 N_c m_Q } \right)^{1/2}
\mathcal{B}_{\rho\sigma} \; \left[ \frac{1}{2}( I^{\rho \mu}
I^{\sigma \nu} + I^{\sigma \mu} I^{\rho \nu})
    -\frac{1}{3} I^{\rho\sigma} I^{\mu \nu} \right]
\epsilon_{\mu \nu}(\lambda)\ . \label{M-3P2}
\end{eqnarray}
\label{M-3}
\end{subequations}
Here, $\langle O_1 \rangle_{\chi_{QJ}}$ are the NRQCD matrix
elements; the tensor $I^{\mu \nu}$ reads
\bqa I^{\mu \nu} = -g^{\mu \nu} + \frac{P^\mu P^\nu}{4 m_Q^2}\ ,
\eqa
and the sums of various polarizations of spin-1 polarization vector
and spin-2 polarization tensors give
\begin{subequations}
\begin{eqnarray}
\sum_{\lambda}
\epsilon^{\mu}(\lambda)\epsilon^{\nu}(\lambda)&=&I^{\mu\nu},
\\
\sum_{\lambda}\epsilon^{\mu\nu}(\lambda)
\epsilon^{\rho\sigma}(\lambda)&=&
\frac{1}{2}(I^{\mu\rho}I^{\nu\sigma} + I^{\mu\sigma}I^{\nu\rho}) -
\frac{1}{3}I^{\mu\nu}I^{\rho\sigma}\ .
\end{eqnarray}
\end{subequations}

\subsection{The light-cone distribution amplitudes}

To calculate the hadron's distribution amplitude, the bi-spinors
$v\bar{u}$ of the quark-antiquark pairs of final state mesons should
be replaced by corresponding light-cone projectors. The light-cone
projectors for light mesons can be readily obtained from Ref.
\cite{beneke}, which is slightly different from what we used in the
following by a factor of $(-i)$ in convention. The two-particle
light-cone projection operator for a light pseudoscalar in momentum
space up to twist-3 reads
\begin{equation}
M^P = \frac{- \, f_P}{4}
  \, \Bigg\{ \not\!p' \gamma_5 \, \phi(u)  -
   \mu_P \gamma_5 \left(  \phi_\rho(u)
       - i \sigma_{\mu\nu}\, n_-^\mu n_+^\nu \,
       \frac{\phi^\prime_\sigma(u)}{12}
 +i \sigma_{\mu\nu}  p'{}^\mu \, \frac{\phi_\sigma(u)}{6}
       \, \frac{\partial}{\partial
         k_\perp{}_{\nu}} \right)
 \Bigg\}\ .
\label{pimeson2}
\end{equation}
Here,  $p'$ stands for the momentum of meson; the parameters $\mu_P$
are $m_\pi^2/(m_u + m_d)$, $m_K^2/(m_u + m_s)$ and
$m_{\eta_c}^2/2m_c^{\overline{MS}}$ in the $\overline{MS}$ scheme
for pion, kion and $\eta_c$, respectively; $\phi_\rho(u)$ and
$\phi_\sigma$ are twist-3 distribution amplitudes; $n_-$ and $n_+ $
are light-like vectors satisfying $n_-^2=0,\ n_+^2=0,\ n_-\cdot
n_+=2$; the component quarks' momenta are assigned as
\bqa && k_1^\mu = u E n_-^\mu + k_\perp^\mu + \frac{\vec
k_\perp^2}{4uE} \, n_+^\mu , \qquad k_2^\mu = \bar u E n_-^\mu -
k_\perp^\mu + \frac{\vec k_\perp^2}{4\bar uE} \, n_+^\mu\ .
\eqa
Note that the term which involves the transverse momentum derivative
acts on the hard scatter amplitude before collinear limit $k_1=u
p\;\! '=u E n_-$ is taken. The asymptotic limit of the leading twist
distribution amplitude takes the form $\phi(u)=6 u \bar{u}$. The two
twist-3 distribution amplitudes $\phi_\rho$ and $\phi_\sigma$ can be
obtained by solving the equations of motion and are determined as
$\phi_\rho(u)=1$ and $\phi_\sigma(u)=6 u \bar{u}$.

In the momentum space, the transverse and longitudinal light-cone
projections of vector meson reads
\bqa
M^V_\parallel &=& \frac{f_V}{4} \, \frac{m_V(\varepsilon\cdot
  n_+)}{2 E} \,E\,
 \slashchar{n}_- \,\phi_\parallel(u)
 + \frac{f_{V}^{T} m_V}{4}  \,\frac{m_V(\varepsilon\cdot n_+)}{2 E}
 \, \Bigg\{-\frac{i}{2}\,\sigma_{\mu\nu} \,  n_-^\mu  n_+^\nu \,
 h_\parallel^{(t)}(u)
\nonumber\\[0.1cm]
&& \hspace*{-0.0cm} - \,i E\, \int_0^u dv \,(\phi_\perp(v) -
h_\parallel^{(t)}(v)) \
     \sigma_{\mu\nu}   n_-^\mu
     \, \frac{\partial}{\partial k_\perp{}_\nu}
  +(1 - \frac{f_V (m_1+m_2)}{f_{V}^{T} m_V})
  \frac{h_\parallel'{}^{(s)}(u)}{2}\Bigg\}\, \Bigg|_{k=u
  p'}
\eqa and \bqa M^V_\perp &=& \frac{1}{4}f^T_V
E\slashchar{\varepsilon}_\perp \slashchar{n}_-
\phi_\perp(u)+\frac{1}{4}f_Vm_V\left[\slashchar{\varepsilon}_\perp
g_\perp^{(v)}(u)-E\slashchar{n}_-
\int^u_0\!dv\left(\phi_\parallel(v)-g_\perp^{(v)}(v)\right)
\varepsilon_\perp^\sigma\frac{\partial}{\partial
k_\perp^\sigma}\right]\nonumber\\
&+&\frac{i}{4}\left(f_{V} - f_{V}^{T} \frac{m_{1} +
m_{2}}{m_{V}}\right) m_{V} \varepsilon_{\mu\nu\rho \sigma}
\varepsilon_\perp^{\nu} n_-^{\rho}\gamma^\mu\gamma_5
\left(n_+^\sigma\frac{g'^{(a)}_\perp(u)}{8}-E\frac{
g^{(a)}_\perp(u)}{4}\frac{\partial}{\partial
k_{\perp\sigma}}\right)\Bigg{|}_{k=up'}\ ,
\eqa
respectively. Here, the transverse polarization vector
\begin{equation}
\varepsilon_\perp^\mu \equiv \varepsilon^\mu -
\frac{\varepsilon\cdot n_+}{2}\,n_-^\mu- \frac{\varepsilon\cdot
n_-}{2}\,n_+^\mu\ .
\end{equation}
In the CMS of the initial quarkonium state, the light-cone vectors
of final state mesons can be chosen as
\bqa n_-=(1,\ \sin\theta,\ 0,\ \cos\theta)\ ,\ \ \ \ \ n_+=(1,\
-\sin\theta,\ 0,\ -\cos\theta)\ .
\eqa
For the vector meson moving along $n_-$, its polarization vectors
are
\bqa \varepsilon(+)& = & (0\ , -\frac{\cos\theta}{\sqrt{2}}\ ,
-\frac{i}{\sqrt{2}}\ , \frac{\sin\theta}{\sqrt{2}})\
,\nonumber\\\varepsilon(-)& = &(0, \frac{\cos\theta}{\sqrt{2}}\ ,
-\frac{i}{\sqrt{2}}\ , -\frac{\sin\theta}{\sqrt{2}})\ ,\nonumber\\
\varepsilon(0)& = &(\frac{\sqrt{E^2-m_V^2}}{m_V}\
,\frac{E\sin\theta}{m_V}\ ,\ 0\ , \frac{E\cos\theta}{m_V})\ . \eqa

The higher twist LCDAs are related to the leading twist ones through
Wandzura-Wilczek relations  \cite{qsr1}
\bqa
  g_\perp^{(v)}(u) &=& \frac12 \left[ \,\int_0^u
    \frac{\phi_\parallel(v)}{\bar v} \, dv  + \int_u^1
    \frac{\phi_\parallel(v)}{v}\, dv  \right] ,
\\
  g_\perp^{(a)}(u) &=& 2 \left[ \bar u \int_0^u
    \frac{\phi_\parallel(v)}{\bar v}\, dv  + u \, \int_u^1
    \frac{\phi_\parallel(v)}{v}\, dv  \right],
\\
    h_\parallel^{(t)}(u) &=& (2u-1) \left[ \,\int_0^u
    \frac{\phi_\perp(v)}{\bar v} \, dv  - \int_u^1
    \frac{\phi_\perp(v)}{v}\, dv  \right] ,
\\
  h_\parallel^{(s)}(u) &=& 2 \left[ \bar u \int_0^u
    \frac{\phi_\perp(v)}{\bar v}\, dv  + u \, \int_u^1
    \frac{\phi_\perp(v)}{v}\, dv  \right].
\label{wwrela} \eqa

Generally, while the above distribution amplitudes are convoluted
with the hard part of a specific process, the troublesome endpoint
divergences appears \cite{BBNS,BRY,Cheng}. In this work the
logarithmic and linear infrared divergences, appearing when the
convolution integrations are performed, are attributed to certain
complex quantities similar as the prescription used in Refs.
\cite{BBNS,BRY,Cheng}, as
\begin{equation}
  \label{eq:XA}
  \int_0^1 \frac{d u}{u} \to X_a\ ,
  \qquad\qquad
  \int_0^1 d u\,\frac{\ln u}{u} \to -\frac{1}{2} \left(X_a\right)^2\ ,
  \qquad\qquad
  \int_0^1 \frac{d u}{u^2} \to X_l-1\ .
\end{equation}
This measure is different from what in the pQCD approach taken in
Ref.\cite{zhou}, where the Sudakov form factor is introduced in the
convolution integration. Here, the first and the second are just the
same as given in Ref. \cite{BRY}, and the third one should be less
than $1$ in magnitude, which is smaller than what in Ref. \cite{BRY}
because here the integral does not divergent at $u=1$. We take the
same assumption about $X_a$ and $X_l$ as in Ref. \cite{BRY}, i.e.,
they are universal to all final states with magnitudes around
$\ln(m_Q/\Lambda_{QCD})$ and $m_Q/\Lambda_{QCD}$, respectively. And,
hence they can be constructed as \bqa X_a = (1 + \rho_a e^{i
\phi_a})\ln(\frac{m_Q}{\Lambda_{QCD}}), \ X_l=(1+\rho_l e^{i
\phi_l})\frac{m_Q}{\Lambda_{QCD}}\ . \eqa
The values of $\rho$ and $\phi$ are such constrained that $\rho \leq
1$ and $\phi$ is arbitrary. In this work we take $\rho_a = \rho_l =
0.5$ and $ \phi_a = \phi_l = 90^\circ$ by default, which is slightly
different from what used in Refs. \cite{BRY,Cheng} in $B$ and $B_s$
decays. The asymptotic amplitudes $\phi(u) = \phi_\sigma(u) =
\phi_\perp(u) = \phi_\parallel(u)=6 u \bar u$ are taken in the
calculation. To test the parameter dependence, we let $\rho_l$ and
$\phi_l$ to vary the same way as $\rho_a$ and $\phi_a$,
respectively. Note that all these procedures in dealing with the
divergences lead the calculation results to be model dependant.

\subsection{The decay width}

Once one knows the distribution amplitude, the decay widths can be
readily obtained. The general polarized decay width reads:
\bqa
\Gamma_{\lambda_1,\lambda_2}^J&=&\frac{1}{2M_{\chi_{QJ}}8\pi}\frac{1}{2J+1}
\sqrt{1-\frac{4m^2}{M_{\chi_{QJ}}^2}}\int_{-1}^{1}\frac{d
\cos\theta}{2}|\mathcal {M}_{\lambda_1,\lambda_2}^J|^2\ .
\label{width} \eqa
Here, $m$ is the mass of light meson and $J$ stands for the quantum
number of total angular momentum. In case the final states are
identical mesons, a statistical factor of $1/2$ should be added to
the width (\ref{width}). While final states being vector mesons, the
total width then is:
\bqa \Gamma^J_{total}
&&=\Gamma^J_{0,0}+2\Gamma^J_{+,+}+2\Gamma^J_{+,-}+4\Gamma^J_{0,+}\ .
\eqa

\subsection{The calculation procedure}
In our calculation, the computer algebra system {\bf MATHEMATICA} is
employed with the help of the packages {\bf FEYNCALC}
\cite{Feyncalc}. {\bf FEYNCALC} is used to trace the Dirac matrices.
The derivatives respect to the momentum are performed by means of
the {\bf MATHEMATICA} code developed by ourself, and it has been
tested before applying to the real calculation by recalculating the
photonic decay rates of the $\chi_{c0}$ and $\chi_{c2}$ at the
leading order and getting an agreement with Ref.
\cite{Braaten:2002fi}. The final analytical results are a little bit
too lengthy to be listed here. The analytical expressions for
$\mathcal{B}_{\rho\sigma}$ in equation (4) for various processes
with different final states (VV or PP) are presented in the appendix
for the sake of the reader¡¯s convenience while they use.

\section{Numerical calculation and discussion}

\subsection{Input parameters}

Before carrying out numerical calculations, the input parameters
need to be fixed. As for the NRQCD matrix elements, we take those
used in Refs. \cite{jungil Lee,bodwin:0710hg,bodwin:2006,chyq},
\begin{eqnarray}\label{numericalmatrix}
 \frac{1}{3}\vert\langle
\chi_{c0}|\psi^\dagger(-\frac{i}{2}\tensor{{\bf D}}\cdot{\bf
\sigma})\chi|0\rangle\vert^2 &=&0.051\ \textrm{GeV}^5 \ ,\nonumber\\
\frac{1}{2}\vert\langle
\chi_{c1}|\psi^\dagger(-\frac{i}{2}\tensor{{\bf
D}}\times{\bf\sigma}\cdot{\bf\epsilon}_H)
\chi|0\rangle\vert^2 &=&0.060\ \textrm{GeV}^5\ ,\nonumber\\
\vert\sum_{ij}\langle
\chi_{c2}|\psi^\dagger(-\frac{i}{2}
\tensor{D}^{(i}\sigma^{j)}\epsilon^{ij}_H)\chi|0\rangle\vert^2
&=&0.068\ \textrm{GeV}^5\ , \nonumber\\
\frac{1}{3}\vert\langle
\chi_{b0}|\psi^\dagger(-\frac{i}{2}\tensor{{\bf D}}\cdot{\bf
\sigma})\chi|0\rangle\vert^2 &=&2.03\ \textrm{GeV}^5\ , \nonumber\\
\frac{1}{2}\vert\langle
\chi_{b1}|\psi^\dagger(-\frac{i}{2}\tensor{{\bf
D}}\times{\bf\sigma}\cdot{\bf\epsilon}_H)
\chi|0\rangle\vert^2 &=&2.03\ \textrm{GeV}^5\ ,\nonumber\\
\vert\sum_{ij}\langle \chi_{b2}|\psi^\dagger(-\frac{i}{2}
\tensor{D}^{(i}\sigma^{j)}\epsilon^{ij}_H)\chi|0\rangle\vert^2
&=&2.03\ \textrm{GeV}^5 \ .
\end{eqnarray}
The leading order running coupling constant
\begin{eqnarray*}
 \alpha_s(\mu) &=& \frac{4\pi}{b_0 \ln(\mu^2/\Lambda_\mathrm{QCD}^2)}
\end{eqnarray*}
is employed with $b_0=\frac{33-2n_f}{3}$. In numerical evaluation,
we take the quark flavor number $n_f=4$ and
$\Lambda_{QCD}=0.225\textrm{GeV}$; the interaction scale $\mu=2m_b$
for the bottomonium decay processes, and $\mu=2m_c$ for the
charmonium decay processes; the up and down quarks are taken to be
massless, while the strange quark mass $m_s = 0.10$GeV, the charm
quark mass $m_c=1.4\pm0.2$GeV in NRQCD calculation, and the bottom
quark mass $m_b=4.8\pm0.1$GeV; the masses of quarkonia are obtained
from the PDG \cite{PDG}, which are $M_{\chi_{b0}}=9.859$GeV,
$M_{\chi_{b1}}=9.892$GeV, $M_{\chi_ {b2}}=9.912$GeV,
$M_{\chi_{c0}}=3.414$GeV, $M_{\chi_{c1}}=3.510$GeV, and $M_{\chi_
{c2}}=3.556$GeV\ .

The input parameters for final state vector and pseudoscalar mesons
are presented in Table \ref{parameter}, which are obtained from
Refs.\cite{PDG,shq,bra2}. For $\eta_c$, the charm quark mass in the
$\overline{MS}$ scheme is taken to be $m_c^{\overline{MS}} = 1.2$GeV
in light cone calculation \cite{shq}.
\begin{table}[h]\caption{Input parameters for final state mesons.}
\label{parameter}
\begin{tabular}{c|c|c|c|c|c}\hline
&$\rho$&$\bar{K}^* $&$\omega
$&$\phi$&$J/\psi$\\
\hline $m_V$[{\rm MeV}]&770&892&782&1020&3097 \\ \hline$f_V[{\rm
MeV}]$ &$205\pm 9$&$217\pm5$&$195\pm3$&$231\pm4$&$416\pm5$\\\hline
$f^T_V[{\rm MeV}]$
&$160\pm10$&$170\pm10$&$145\pm10$&$200\pm10$&$379\pm21$\\
\hline\hline &$\pi^+(\pi^-)$&$K^+(K^-) $&$\eta_c
$\\
\hline $m_P$[{\rm MeV}]&139.6&493.7&2980 \\\hline $f_P[{\rm MeV}]$
&$130.4\pm0.2$&$156.1\pm0.8$&$373\pm64$\\
\hline $\mu_P[{\rm MeV}]$ &$1500$&$1700$&$3700$\\
\hline
\end{tabular}
\end{table}

\subsection{Numerical results and Discussions}

\begin{table}[h]
\caption{\label{table3}%
The polarized decay widths for $\chi_{bJ}\to VV$ in unit of eV.\\}
\begin{ruledtabular}
\begin{tabular}{c|c|c|c|c}
&$\Gamma_{0,\ 0}$[eV]&$
\Gamma_{+,\ +}$[eV]   & $
\Gamma_{+,\ 0}$[eV]   &  $\Gamma_{+,\ -}$[eV] \,\\
\hline
$\chi_{b0}\to\rho^0\rho^0$
&$2.60_{-0.39-0.09}^{+0.48+0.09}$
& $0.009_{-0.002-0.009}^{+0.003+0.079}$ & --- & --- \\
$\chi_{b0}\to\rho^+\rho^-$
&$5.20_{-0.78-0.18}^{+0.96+0.18}$
& $0.018_{-0.004-0.018}^{+0.006+0.158}$ & ---&--- \\
$\chi_{b0}\to K^* \bar{K}^*$
&$6.35_{-0.10-0.14}^{+0.12+0.38}  $
& $0.034_{-0.008-0.034}^{+0.010+0.315}$ & --- & --- \\
$\chi_{b0}\to K^{*+} K^{*-}$
&$6.35_{-0.10-0.14}^{+0.12+0.38}  $
& $0.034_{-0.008-0.034}^{+0.010+0.315}$ & --- & --- \\
$\chi_{b0}\to\omega \omega$
&$2.13_{-0.32-0.07}^{+0.39+0.07}  $
& $0.007_{-0.002-0.007}^{+0.002+0.068}$ & --- & --- \\
$\chi_{b0}\to \phi \phi$
&$3.93_{-0.59-0.15}^{+0.71+0.41}  $
& $0.034_{-0.007-0.034}^{+0.010+0.306}$ & --- & --- \\
$\chi_{b0}\to J/\psi J/\psi$
&$10.76_{-1.10-6.79}^{+1.21+21.35}$
& $11.15_{-2.46-10.89}^{+3.22+93.96}  $ & --- & --- \\
\hline
$\chi_{b1}\to\rho^0 \rho^0$
& --- & --- & $1.2_{-0.2-1.2}^{+0.3+25.9}\times10^{-3}$ & --- \\
$\chi_{b1}\to\rho^+ \rho^-$
& --- & --- & $2.4_{-0.4-2.4}^{+0.6+51.8}\times10^{-3}$ & --- \\
$\chi_{b1}\to K^* \bar{K}^*$
& --- & --- & $5.9_{-1.1-5.9}^{+1.4+124.6}\times10^{-3}$ & --- \\
$\chi_{b1}\to K^{*+} K^{*-}$
& --- & --- & $5.9_{-1.1-5.9}^{+1.4+124.6}\times10^{-3}$ & --- \\
$\chi_{b1}\to \omega \omega$
& --- & --- & $1.2_{-0.2-1.1}^{+0.3+24.9}\times10^{-3}$ & --- \\
$\chi_{b1}\to \phi \phi$
& --- & --- & $3.3_{-0.6-3.3}^{+0.8+75.7}\times10^{-3}$ & --- \\
$\chi_{b1}\to J/\psi J/\psi$
& --- & --- & $0.51_{-0.09-0.25}^{+0.11+10.58}$ & --- \\
\hline
$\chi_{b2}\to\rho^0 \rho^0$
& $0.14_{-0.02-0.03}^{+0.03+0.03}$
& $0.50_{-0.11-0.41}^{+0.14+6.80}\times10^{-3} $
& $0.024_{-0.005-0.023}^{+0.006+0.145}$
& $0.61_{-0.10-0.10}^{+0.12+0.17}$ \\
$\chi_{b2}\to\rho^+ \rho^-$
& $0.28_{-0.04-0.06}^{+0.06+0.06}$
& $1.0_{-0.22-0.82}^{+0.28+13.60}\times10^{-3} $
& $0.048_{-0.010-0.046}^{+0.012+0.290}$
& $1.22_{-0.20-0.20}^{+0.24+0.34}$ \\
$\chi_{b2}\to K^* \bar{K}^*$
& $0.36_{-0.05-0.09}^{+0.07+0.10}$
& $2.24_{-0.50-1.80}^{+0.65+30.63}\times10^{-3}$
& $0.082_{-0.015-0.077}^{+0.019+0.485}$
& $1.65_{-0.26-0.34}^{+0.33+0.59}$\\
$\chi_{b2}\to K^{*+} K^{*-}$
& $0.36_{-0.05-0.09}^{+0.07+0.10}$
& $2.24_{-0.50-1.80}^{+0.65+30.63}\times10^{-3}$
& $0.082_{-0.015-0.077}^{+0.019+0.485}$
& $1.65_{-0.26-0.34}^{+0.33+0.59}$\\
$\chi_{b2}\to \omega \omega$
& $0.18_{-0.03-0.03}^{+0.03+0.04}$
& $0.66_{-0.15-0.59}^{+0.19+9.1}\times10^{-3}  $
& $0.032_{-0.006-0.030}^{+0.007+0.188}$
& $0.78_{-0.12-0.13}^{+0.15+0.22}$ \\
$\chi_{b2}\to\phi \phi$
& $0.24_{-0.04-0.09}^{+0.04+0.09}$
& $2.78_{-0.61-2.71}^{+0.80+30.10}\times10^{-3}$
& $0.077_{-0.014-0.076}^{+0.018+0.474}$
& $1.57_{-0.25-0.32}^{+0.31+0.65}$\\
$\chi_{b2}\to J/\psi J/\psi$
& $3.81_{-0.63-3.81}^{+0.77+4.84}$
& $2.52_{-0.56-2.14}^{+0.73+13.86}             $
& $5.46_{-0.97-4.94}^{+1.19+36.39}    $
& $51.91_{-9.85-33.66}^{+12.46+140.26}$ \\
\end{tabular}
\end{ruledtabular}
\end{table}
By virtue of the formulae provided in section~II and combined with
tensors given in appendix, one can readily get the analytical decay
widths for final state mesons with specified helicity. After
substituting the input parameters in preceding section to the
analytical expressions, the numerical results are obtained. The
magnitudes of decay widths for various processes, including helicity
decay widths for final state vector mesons, are presented in
Tables~II to IV.

In our calculation, the uncertainties are estimated as follows. The
first one comes from the uncertainty of heavy quark mass $m_Q$. The
uncertainty in $\alpha_s(2m_Q)$ induced by $m_Q$ has also been taken
into account in the evaluation. The second one comes from the
uncertainties of parameters $\rho$ and $\phi$, and is evaluated by
taking the decay width as function of $\rho$ and $\phi$. These two
sources of uncertainty are the major ones in the calculation of this
work.

\begin{table}[h]
\caption{\label{table4}%
The polarized decay widths for $\chi_{cJ}\to VV$ in unit of keV.}
\begin{ruledtabular}
\begin{tabular}{c|c|c|c|c}
&$\Gamma_{0,\ 0}$[keV] & $\Gamma_{+,\ +}$[keV]
& $\Gamma_{+,\ 0}$[keV]   &  $\Gamma_{+,\ -}$[keV] \,\\
\hline $\chi_{c0}\to\rho^0\rho^0$& $3.96_{-2.62-0.90}^{+9.63+0.99}
$
& $0.31_{-0.25-0.31}^{+1.84+2.81}$ & --- & --- \\
$\chi_{c0}\to\rho^+\rho^-$
&$7.92_{-5.24-1.80}^{+19.26+1.98} $
& $0.62_{-0.50-0.62}^{+3.68+5.62}$   & --- & --- \\
$\chi_{c0}\to K^* \bar{K}^*$
&$8.67_{-5.66-2.56}^{+19.17+2.89} $
& $1.32_{-1.07-1.32}^{+7.93+12.16}$ & --- & --- \\
$\chi_{c0}\to K^{*+} K^{*-}$
&$8.67_{-5.66-2.56}^{+19.17+2.89} $
& $1.32_{-1.07-1.32}^{+7.93+12.16}$ & --- & --- \\
$\chi_{c0}\to\omega \omega$
&$3.16_{-2.08-0.68}^{+7.58+0.74}  $
& $0.26_{-0.22-0.26}^{+1.59+2.71}$ & --- & --- \\
$\chi_{c0}\to \phi \phi$
&$4.21_{-2.64-0.71}^{+7.27+2.36}  $
& $1.09_{-0.89-1.08}^{+6.56+9.95}$ & --- & --- \\
\hline
$\chi_{c1}\to\rho^0 \rho^0$
& --- & --- & $0.037_{-0.028-0.030}^{+0.14+0.16}$  & --- \\
$\chi_{c1}\to\rho^+ \rho^-$
& --- & --- & $0.074_{-0.056-0.060}^{+0.28+0.32}    $  & --- \\
$\chi_{c1}\to K^* \bar{K}^*$
& --- & --- & $0.097_{-0.72-0.074}^{+0.35+0.43}   $  & --- \\
$\chi_{c1}\to K^{*+} K^{*-}$
& --- & --- & $0.097_{-0.72-0.074}^{+0.35+0.43}   $  & --- \\
$\chi_{c1}\to \omega \omega$
& --- & --- & $0.024_{-0.017-0.020}^{+0.089+0.102}    $  & --- \\
$\chi_{c1}\to \phi \phi$
& --- & --- & $0.110_{-0.081-0.088}^{+0.358+0.476}    $  & --- \\
\hline
$\chi_{c2}\to\rho^0 \rho^0$
& $0.56_{-0.39-0.41}^{+1.79+0.59}$
& $0.029_{-0.024-0.022}^{+0.174+0.271}$
&$0.11_{-0.08-0.11}^{+0.42+0.65}$
& $2.93_{-2.15-1.32}^{+11.42+2.62}$ \\
$\chi_{c2}\to\rho^+ \rho^-$
& $1.12_{-0.78-0.82}^{+3.58+1.18}$
& $0.058_{-0.048-0.044}^{+0.348+0.542}$
&$0.22_{-0.16-0.22}^{+0.84+1.30}$
& $5.86_{-4.30-2.64}^{+22.84+5.24}$ \\
$\chi_{c2}\to K^* \bar{K}^*$
& $1.47_{-1.04-1.24}^{+4.43+1.76}$
& $0.138_{-0.112-0.108}^{+0.833+1.109}$
& $0.34_{-0.25-0.25}^{+1.20+2.00}$
& $8.95_{-6.66-4.49}^{+36.90+10.21}$\\
$\chi_{c2}\to K^{*+} K^{*-}$
& $1.47_{-1.04-1.24}^{+4.43+1.76}$
& $0.138_{-0.112-0.108}^{+0.833+1.109}$
& $0.34_{-0.25-0.25}^{+1.20+2.00}$
& $8.95_{-6.66-4.49}^{+36.90+10.21}$\\
$\chi_{c2}\to \omega \omega$
& $0.72_{-0.51-0.53}^{+2.27+0.77}$
& $0.039_{-0.031-0.022}^{+0.232+0.360} $
& $0.14_{-0.10-0.10}^{+0.53+0.83}$
& $3.78_{-2.78-1.74}^{+14.8+3.45}$ \\
$\chi_{c2}\to\phi \phi$
& $1.15_{-0.81-1.10}^{+3.10+1.70}$
& $0.156_{-0.126-0.147}^{+0.938+1.090}$
& $0.25_{-0.18-0.24}^{+0.80+1.55}$
& $8.52_{-6.35-4.21}^{+35.46+10.58}$\\
\end{tabular}
\end{ruledtabular}
\end{table}

\begin{table}[here]
\caption{\label{table5}%
The decay widths of $\chi_{QJ}\to PP$ processes.}
\begin{ruledtabular}
\begin{tabular}{ccccc}
  &  & $\eta_c\eta_c$ & $K^+K^-$   & $\pi^+\pi^-$\,\\
\hline \multirow{2}{*}{$\Gamma(\chi_{bJ}) $[eV]}&
$\chi_{b0}$     & $20.90_{-3.11-3.24}^{+3.72+3.44} $
& $1.75_{-0.27-0.06}^{+0.31+0.06}$
& $0.85_{-0.13-0.02}^{+0.16+0.02}$ \\
&$\chi_{b2}$     & $4.17_{-0.62-0.65}^{+0.74+0.69} $
& $0.35_{-0.05-0.01}^{+0.06+0.01}$
& $0.17_{-0.02-0.00}^{+0.03+0.00}$ \\
\hline \multirow{2}{*}{$\Gamma(\chi_{cJ}) $[keV]}&
$\chi_{c0}$      &   & $2.95_{-1.97-0.70}^{+7.59+0.77}$
 & $1.53_{-1.03-0.28}^{+4.01+0.30}$ \\
&$\chi_{c2}$     &   & $0.76_{-0.51-0.18}^{+1.95+0.20}$
 & $0.39_{-0.26-0.07}^{+1.03+0.08}$ \\
\end{tabular}
\end{ruledtabular}
\end{table}
The polarized decay widths $\Gamma_{+, +}$ and $\Gamma_{+, 0}$ come
solely from the twist-3 contribution, which therefore do not exist
in Ref.\cite{bra2} where only the leading twist distribution is
considered. The twist-3 is the leading contribution for
$\chi_{Q1}\to VV$ processes. In the literature, the polarized decay
width $\Gamma_{+, 0}$ of $\chi_{Q1}$ from the contribution of QED
correction had been taken into account \cite{zdf}, however we find
the QED contribution is several orders less than what from QCD, and
hence are negligible. From Tables \ref{table3} and \ref{table4} we
notice that the twist-3 contributions $\Gamma_{+, +}$ and
$\Gamma_{+, 0}$ are prominent for polarized widths of
$\chi_{b0}/\chi_{b2}\to J/\psi J/\psi$ processes. While for
processes $\chi_{c0}/\chi_{c2}\to K^*\bar{K}^*$ and
$\chi_{c0}/\chi_{c2}\to \phi\phi$, though the twist-3 contributions
are not predominant, they are still important.

Notice that the higher twist contributions are suppressed by powers
of $\frac{m_V}{m_Q}$, and in dealing with the infrared divergences
appearing in the next-to-leading contributions in twist expansion,
factors of $\ln(\frac{m_Q}{\Lambda_{QCD}})$,
$\ln^2(\frac{m_Q}{\Lambda_{QCD}})$, or $\frac{m_Q}{\Lambda_{QCD}}$
are induced via the regularization procedure (\ref{eq:XA}). Except
for polarized decay width $\Gamma_{+,\ 0}$, which comes from the
interference of twist-2 and -3 terms and hence has the power
suppression of $\frac{m_V}{m_Q}$ at the amplitude level, other
polarized decay widths from twist-3 contributions all have power
suppression of $(\frac{m_V}{m_Q})^2$ at the amplitude level as
expected. The $\Gamma_{0,\ 0}$ contains only
$\ln(\frac{m_Q}{\Lambda_{QCD}})$ form, the $\Gamma_{+,\ 0}$ contains
both $\ln(\frac{m_Q}{\Lambda_{QCD}})$ and
$\ln^2(\frac{m_Q}{\Lambda_{QCD}})$ forms, while $\Gamma_{+,\ +}$ and
$\Gamma_{+,\ -}$ contain all three kinds of regularized forms. So,
as $m_V\sim \Lambda_{QCD}$, the convergence in twist expansion is
hold well by power counting.

Of the processes $\chi_{b0}/\chi_{b2}\to J/\psi J/\psi$, the twist
expansion fails. For $\chi_{b0}\to J/\psi J/\psi$, the polarized
decay width $\Gamma_{+,\ +}$, which comes merely from the twist-3
contribution, is even bigger than $\Gamma_{0,\ 0}$, mainly the
twist-2 contribution. For $\chi_{b2}\to J/\psi J/\psi$ process, the
twist-3 contribution $\Gamma_{+,\ 0}$ is also larger than
$\Gamma_{0,\ 0}$. This is not a big surprise, though $m_{J/\psi}$ is
smaller than $m_{\chi_{bJ}}$, it is still much larger than
$\Lambda_{QCD}$ in comparison with other light mesons, which spoils
the twist expansions. Hence, the perturbative calculation on
$\chi_{bJ}\to J/\psi J/\psi$ in this work can be treated as a
qualitative estimation.

To estimate the branching ratios we use the following expressions
for the total bottomonium decay widths  \cite{bra2,ttw}: \bqa
\Gamma_{\chi_{b0}}&&=
    \frac{3C_F}{N_c}\pi\alpha_s^2
    \frac{\langle O_P^{bb}\rangle}{m_b^4} +
    \frac{n_f}{3}\pi\alpha_s^2 \frac{\langle O_8\rangle}{m_b^2}
    = 0.68\ \mathrm{MeV},\\
\Gamma_{\chi_{b2}}&&=
    \frac{4C_F}{5N_c}\pi\alpha_s^2
    \frac{\langle O_P^{bb}\rangle}{m_b^4} +
    \frac{n_f}{3}\pi\alpha_s^2 \frac{\langle O_8\rangle}{m_b^2}
    = 0.20\ \mathrm{MeV},\\
\Gamma{\chi_{b1}}&&=\frac{C_F\alpha_s^3}{N_c}\left[
        \left(\frac{587}{54}-\frac{317}{288}\pi^2\right)C_A+
        \left(-\frac{16}{27}-\frac{4}{9}\ln\frac{\Lambda}{2m_b}\right)n_f
\right]\frac{\langle O_P^{bb}\rangle}{m_b^4}
    +\frac{n_f}{3}\pi\alpha_s^2 \frac{\langle O_8\rangle}{m_b^2}\nonumber \\
&&= 0.09\ \mathrm{MeV}.
\eqa The total decay widths for $\chi_{cJ}$ are obtained from PDG
\cite{PDG}, i.e., $\Gamma_{\chi_{c0}}=10.3\mathrm{MeV},\
\Gamma_{\chi_{c1}}=0.86\mathrm{MeV},$ and $
\Gamma_{\chi_{c2}}=1.97\mathrm{MeV}.$

The branching ratios of different processes are listed in Table V.
They are evaluated by taking the center value of each unpolarized
width and divided by the total decay widths of $\chi_{QJ}$.

\begin{table}[here]
\caption{\label{table6}%
The branching fractions of various processes of p-wave heavy
quarkonium exclusive decays to double mesons.}
\begin{ruledtabular}
\begin{tabular}{ccccccc}
&$\rho^0 \rho^0$ & $K^* \bar{K}^*$
& $\omega \omega$& $\phi \phi$ & $J/\psi J/\psi$ \,\\
\hline \multirow{1}{*}{$\chi_{b0}$}&
    $3.9\times10^{-6}$ & $9.3\times10^{-6}$
    & $3.2\times10^{-6}$ & $5.9\times10^{-6}$ & $4.9\times10^{-5}$\\
 \multirow{1}{*}{$\chi_{b1}$}&
   $4.8\times10^{-8}$ & $2.4\times10^{-7}$
   & $4.8\times10^{-8}$ & $1.3\times10^{-7}$ & $2.1\times10^{-5}$\\
 \multirow{1}{*}{$\chi_{b2}$}
    & $7.3\times10^{-6}$ & $2.0\times10^{-5}$
    & $9.3\times10^{-6}$ & $1.8\times10^{-5}$ & $6.7\times10^{-4}$ \\
\hline \multirow{1}{*}{$\chi_{c0}$}&
   $4.4\times10^{-4}$ & $10.9\times10^{-4}$
   & $3.6\times10^{-4}$ & $6.2\times10^{-4}$ \\
 \multirow{1}{*}{$\chi_{c1}$}&
   $1.7\times10^{-4}$ & $4.5\times10^{-4}$
   & $1.1\times10^{-4}$ & $5.1\times10^{-4}$ \\
 \multirow{1}{*}{$\chi_{c2}$}
& $35\times10^{-4}$ & $11\times10^{-3}$
& $45\times10^{-4}$ & $98\times10^{-4}$ \\
\hline \multirow{2} {*}
&$\eta_c\eta_c$ & $K^+K^-$ & $\pi^+\pi^-$ \,\\
\hline\multirow{1}{*}{$\chi_{b0}$}&
   $3.1\times10^{-5}$ & $2.6\times10^{-6}$ & $1.3\times10^{-6}$ \\
\multirow{1}{*}{$\chi_{b2}$}&
   $2.1\times10^{-5}$ & $1.8\times10^{-6}$ & $8.5\times10^{-7}$  \\
\hline \multirow{1}{*}{$\chi_{c0}$}
&  & $2.9\times10^{-4}$ & $1.5\times10^{-4}$ \\
 \multirow{1}{*}{$\chi_{c2}$}
&  & $3.8\times10^{-4}$ & $1.9\times10^{-4}$ \\
\end{tabular}
\end{ruledtabular}
\end{table}
\newpage
For comparison, in Table \ref{table7} we also present the total
decay width of $\chi_{bJ}\to J/\psi J/\psi$ obtained in other
calculations. The second and third columns give what we obtained in
this work. The fourth column shows the NRQCD results of
Ref.\cite{zdf}, including QED and relativistic corrections. The last
column presents the calculation results in light-cone formalism up
to twist-2 \cite{bra2}.

\begin{table}[here]
\caption{\label{table7}%
The total width of $\chi_{bJ}\to J/\psi J/\psi$ process in various
calculation methods.}
\begin{ruledtabular}
\begin{tabular}{cccccccc}
[eV]&$\Gamma$[twist-2]& $\Gamma$[twist-3] & $\Gamma$-NRQCD\cite{zdf}
& $\Gamma$-LC twist-2\cite{bra2} &  & \,\\
\hline \multirow{1}{*}{$\chi_{b0}\to J/\psi J/\psi$}
&$21.69_{-2.86}^{+3.32}$ &$33.05_{-6.02-28.48}^{+7.65+209.27}$ &
$5.54$
& $79.\pm3.1\pm25.\pm31.$ & & \\
\multirow{1}{*}{$\chi_{b1}\to J/\psi J/\psi$}
& 0 & $2.04_{-0.36-1.00}^{+0.44+42.32}$ & $9.04\times10^{-7}$ & --- & & & \\
\multirow{1}{*}{$\chi_{b2}\to J/\psi J/\psi$}
&$25.84_{-3.93}^{+4.72}$ & $134.51_{-25.33-95.17}^{+31.91+458.64}$
& $10.6$ & $270.\pm41.\pm93.\pm110.$ & &  \\
\end{tabular}
\end{ruledtabular}
\end{table}

From the results in Table \ref{table7} we notice that the light cone
formalism generally yields more than what from the NRQCD calculation
for $J/\psi$ production, and the results obtained in Ref.\cite{zdf}
are much smaller than ours, especially for the $\chi_{b1}\to J/\psi
J/\psi$ process. Further experimental measurement on $\chi_{b1}\to
J/\psi J/\psi$ process may provide test about these two mechanisms.
On the other hand, within the light cone formalism the results
obtain in Ref. \cite{bra2} are much larger than ours. We have
compared the analytical result of the twist-2 part, find that there
is a redundant factor of two in Ref. \cite{bra2}. Furthermore, the
different choices of input parameters may also induce certain
uncertainties.

\begin{table}[here]
\caption{\label{table8}%
Branch ratios of $\chi_{cJ}\to VV$ and $\chi_{cJ}\to PP$, from
BESIII experiment \cite{bes}, the PDG data \cite{PDG}, and our
calculation.}
\begin{ruledtabular}
\begin{tabular}{ccccccc}
& Br(twist-3) & BESIII \cite{bes} & PDG \cite{PDG} & \,\\
\hline \multirow{1}{*}{$\chi_{c0}\to \phi\phi$}
& $(6.2^{+19.8+21.6}_{-4.3-2.78})\times10^{-4}$  &
$(8.0\pm0.3\pm0.8)\times10^{-4}$ & $(9.2\pm1.9)\times10^{-4}$  \\
\multirow{1}{*}{$\chi_{c1}\to \phi\phi$}
& $(5.1^{+16.7+22.1}_{-3.8-4.1})\times10^{-4}
$ & $(4.4\pm0.3\pm0.5)\times10^{-4}$ & --- \\
\multirow{1}{*}{$\chi_{c2}\to \phi\phi$}
& $(98^{+402+159}_{-73-54})\times10^{-4}$ &
$(10.7\pm0.7\pm1.2)\times10^{-4}$  & $(14.8\pm2.8)\times10^{-4}$  \\
\hline \multirow{1}{*}{$\chi_{c0}\to \omega\omega$}
& $(3.6^{+10.4+6.0}_{-2.4-1.2})\times10^{-4}$
& $(9.5\pm0.3\pm1.1)\times10^{-4}$ & $(22\pm7)\times10^{-4}$ \\
\multirow{1}{*}{$\chi_{c1}\to \omega\omega$}
& $(1.1^{+4.1+4.7}_{-0.8-0.9})\times10^{-4}$
& $(6.0\pm0.3\pm0.7)\times10^{-4}$ & --- \\
\multirow{1}{*}{$\chi_{c2}\to \omega\omega$}
& $(45^{+174+59}_{-33-23})\times10^{-4}$ &
$(8.9\pm0.3\pm1.1)\times10^{-4}$ & $(19\pm6)\times10^{-4}$ \\
\hline \multirow{1}{*}{$\chi_{c0}\to K^*\bar{K}^*$}&
$(10.9^{+34.0+26.4}_{-7.6-5.0})\times 10^{-4}$
& ---  & $(17\pm6\pm1)\times 10^{-4}$ \\
\multirow{1}{*}{$\chi_{c0}\to K^+K^-$}&
$(2.9^{+7.7+1.9}_{-0.7-0.7})\times 10^{-4}$
& ---  &$(6.10\pm0.35)\times 10^{-3}$ \\
\multirow{1}{*}{$\chi_{c0}\to \pi\pi$}&
$(1.5^{+3.9+0.3}_{-1.0-0.3})\times 10^{-4}$
& ---  & $(8.4\pm0.4)\times10^{-3}$\\
\hline \multirow{1}{*}{$\chi_{c2}\to K^*\bar{K}^*$}&
$(10.6^{+43.0+16.4}_{-7.9-5.8})\times 10^{-3}$
& ---  & $(2.5\pm0.5)\times10^{-3}$ \\
\multirow{1}{*}{$\chi_{c2}\to K^+K^-$}&
$(3.8^{+9.9+1.0}_{-2.5-0.9})\times 10^{-4}$ & ---  &
$(10.9\pm0.8)\times 10^{-4}$ \\
\multirow{1}{*}{$\chi_{c2}\to \pi\pi$}&
$(2.0^{+5.2+0.4}_{-1.3-0.3})\times 10^{-4}$
& ---  &$(2.39\pm0.14)\times 10^{-3}$ \\
\end{tabular}
\end{ruledtabular}
\end{table}
The comparison of our calculation with experimental results is given
in Table \ref{table8}. We can see from the table that our results in
general agree with the experiment in all $\chi_{cJ}\to VV$ processes
within an order of magnitude, while the NRQCD calculation in
Ref.\cite{zdf} for $\chi_{c1}\to VV$ processes gives results several
orders smaller than the BESIII measurements. Except for
$\chi_{c2}\to K^+K^-$, the calculated $\chi_{c0}/\chi_{c2}$ to
$\pi\pi$ and $K^+K^-$ decay widths are about an order smaller than
the PDG\cite{PDG} data. To be noted that of all our calculation
results, the $\chi_{cJ}\to \phi\phi$ processes possess larger
branching ratios than that of the $\chi_{cJ}\to \omega\omega$
processes, while experimental data do not tell so all the way. This
is an open question leaving for further study, and could be partly
understood as the difference in the selection of nonperturbative
parameters $\rho$ and $\phi$ in various $\chi_{QJ}$ decay processes.

\section{Summary and Conclusions}

In this work, various exclusive processes of p-wave Quarkonium
decays to two mesons are investigated, where the final state vector
mesons in different polarizations are considered separately. For
light mesons, we expand the light cone distribution amplitudes up to
twist-3, which induces certain end-point singularities in the
calculation. To hurdle these singularities, the prescription used in
dealing with the B meson decay problems is employed, in which two
complex parameters are introduced. Though these parameters induce
some uncertainties in our calculation, we believe our result should
be reliable as an order estimation.

The $\chi_{c1}$ and $\chi_{b1}$ to VV decay processes are induced
entirely by the twist-3 contribution in distribution amplitude
expansion, which are firstly estimated in this work. It turns out
that the higher twist contribution is significant and provides a
possible explanation for the observation of hadron helicity
selection rule violated processes $\chi_{c1}\rightarrow \phi\phi$
and $\omega\omega$ observed by the BESIII collaboration in recently.
We also evaluate the $\chi_{b1}\to J/\psi J/\psi$ process and find
that its branching ratio is big enough to be measured at the LHC and
B-factory experiments.

Note that after this work has been done, the Belle Collaboration
releases new measurement results on the p-wave spin-triplet
bottomonium decays to double charmonium \cite{belle}. Their obtained
upper limits of the branch ratios $Br(\chi_{b0}\to J/\psi
J/\psi)<7.1\times 10^{-5}$, $Br(\chi_{b1}\to J/\psi
J/\psi)<2.7\times 10^{-5}$, and $Br(\chi_{b2}\to J/\psi
J/\psi)<4.5\times 10^{-5}$ at the 90\% confidence level are
compatible with our calculation, that is: $Br(\chi_{b0}\to J/\psi
J/\psi)=(0.7\sim35.9)\times 10^{-5}$, $Br(\chi_{b1}\to J/\psi
J/\psi)=(0.7\sim44)\times 10^{-5}$, and $Br(\chi_{b2}\to J/\psi
J/\psi)=(6.9\sim295)\times 10^{-5}$, while taking the uncertainties
in both theory and experiment into account.

\vspace{0.7cm} {\bf Acknowledgments}

This work was supported in part by the National Natural Science
Foundation of China(NSFC) under the grants 10935012, 10821063 and
11175249.

\vspace{4cm}

\newpage
\appendix
\section{tensor structure of the amplitude}

For vector mesons in the final state, the tensor
${\mathcal{B}}^{\mu\nu}$ reads:
\begin{eqnarray}
{\mathcal{B}}^{\mu\nu} &=&
T^{0}(\mathbf{B}^{\mu\nu}+\mathbf{C}^{\mu\nu}+
\mathbf{D}^{\mu\nu}+\mathbf{S}^{\mu\nu})\ , \label{M-1S0}
\end{eqnarray}
\bqa
 \mathbf{B}^{\mu\nu} &&=\frac{ {f_{V}^T}^2 m_{V}^4
\varepsilon_1 \cdot n_+ \varepsilon_2 \cdot n_-}{{m_Q}^4}
\{{B_{k1}}g^{\mu\nu}+B_{k2}(n_{-}^\mu n_{-}^\nu+n_{-}^\mu
n_{-}^\nu)+B_{k3}(n_{-}^\mu n_{+}^\nu + n_{+}^\mu
n_{-}^\nu)\nonumber\\&&+B_{k4}g_{\perp}^{\mu\nu}\}+\frac{{f_{V}^T}^2
m_{V}^2}{m_Q^2}\{B_{k5}[\varepsilon_{\perp
1}^{\mu}(n_{-}^\nu-n_{+}^\nu)\varepsilon_{2}\cdot
n_{-}+\varepsilon_{\perp
2}^{\mu}(n_{+}^\nu-n_{-}^\nu)\varepsilon_{1}\cdot
n_{-}]\nonumber\\&&+B_{k6}(n_{-}^{\mu}\varepsilon_{\perp
1}^{\nu}\varepsilon_{2}\cdot n_{-}+n_{+}^\mu\varepsilon_{\perp
2}^{\nu}\varepsilon_{1}\cdot
n_{+})+B_{k7}(n_{+}^{\mu}\varepsilon_{\perp
1}^{\nu}\varepsilon_{2}\cdot n_{-}+n_{-}^\mu\varepsilon_{\perp
2}^{\nu}\varepsilon_{1}\cdot n_{+})\}\nonumber\\&&+
{f_{V}^T}^2\{B_{k8}(\varepsilon_{\perp1}^\mu\varepsilon_{\perp2}^\nu+
\varepsilon_{\perp1}^\nu
\varepsilon_{\perp2}^\mu)+B_{k9}\varepsilon_{\perp1}
\cdot\varepsilon_{\perp2}\ g_{\perp}^{\mu\nu}\}\ ,\\\nonumber
\\
\mathbf{C}^{\mu\nu} &&= \frac{f_{V}^2m_{V}^2}{m_Q^2}
\{C_{k1}(\varepsilon_{\perp1}^\mu\varepsilon_{\perp2}^\nu
+\varepsilon_{\perp1}^\nu
\varepsilon_{\perp2}^\mu)+C_{k2}(\varepsilon_{\perp1}
\cdot\varepsilon_{\perp2})g^{\mu\nu}\nonumber\\&&+C_{k3}(n_{-}^\mu
n_{-}^\nu+n_{+}^\mu
n_{+}^\nu)\varepsilon_{\perp1}\cdot\varepsilon_{\perp2}
+C_{k4}(n_{+}^\mu n_{-}^\nu+n_{-}^\mu n_{+}^\nu)\varepsilon_{\perp1}
\cdot\varepsilon_{\perp2}\nonumber\\&&+C_{k5}[\varepsilon_{\perp1}
^\mu(n_{-}^{\nu}-n_{+}^{\nu})\varepsilon_{2}\cdot
n_{-}+\varepsilon_{\perp2}^\mu(n_{+}^{\nu}-n_{-}^{\nu})\varepsilon_{1}\cdot
n_{+}]\nonumber\\&&+C_{k6}(n_{-}^\mu \varepsilon_{\perp1}^\nu
\varepsilon_{2}\cdot n_{-}+n_{+}^\mu
\varepsilon_{\perp2}^\nu\varepsilon_{1}\cdot n_{+})+C_{k7}(n_{+}^\mu
\varepsilon_{\perp1}^\nu \varepsilon_{2}\cdot n_{-}+n_{-}^\mu
\varepsilon_{\perp2}^\nu\varepsilon_{1}\cdot n_{+})\nonumber\\&&+
C_{k8}\varepsilon_{1}\cdot n_{+}\varepsilon_{2}\cdot
n_{-}g^{\mu\nu}+C_{k9}\varepsilon_{1}\cdot n_{+}\varepsilon_{2}\cdot
n_{-}(n_{-}^\mu n_{-}^\nu+n_{+}^\mu
n_{+}^\nu)\nonumber\\&&+C_{k10}\varepsilon_{1}\cdot
n_{+}\varepsilon_{2}\cdot n_{-}(n_{+}^\mu n_{-}^\nu+n_{-}^\mu
n_{+}^\nu)\}\ ,\\\nonumber
\\
\mathbf{D}^{\mu\nu} &&= \frac{\tilde{f}_{V}^2
m_{V}^2}{m_Q^2}\{D_{k1}(\varepsilon_{\perp1}^\mu
\varepsilon_{\perp2}^\nu+\varepsilon_{\perp1}
^\nu\varepsilon_{\perp2}^\mu)+D_{k2}
\varepsilon_{\perp1}\cdot\varepsilon_{\perp2}g_{\perp}^
{\mu\nu}+D_{k3}(\varepsilon_{\perp1}
\cdot\varepsilon_{\perp2})g^{\mu\nu}\nonumber\\&&+
D_{k4}\varepsilon_{\perp1}\cdot\varepsilon_{\perp2}(n_{+}^\mu
n_{+}^\nu+n_{-}^\mu
n_{-}^\nu)+D_{k5}\varepsilon_{\perp1}\cdot\varepsilon_{\perp2}(n_{+}^\mu
n_{-}^\nu+n_{-}^\mu n_{+}^\nu)\}\ ,\\\nonumber
\\
\mathbf{S}^{\mu\nu} &&= \frac{-\tilde{f}_{V} f_{V}
m_{V}^2}{m_Q^2}\{S_{k1}(\varepsilon_{\perp1}^\mu
\varepsilon_{\perp2}^\nu+\varepsilon_{\perp1}^\nu
\varepsilon_{\perp2}^\mu)+S_{k2}
(\varepsilon_{\perp1}^\mu\varepsilon_{\perp2}^\nu+\varepsilon_
{\perp1}^\nu\varepsilon_{\perp2}^\mu)
\nonumber\\&&+S_{k3}\varepsilon_{\perp1}\cdot\varepsilon_{\perp2}\
g_{\perp}^{\mu\nu} +S_{k4}\varepsilon_{\perp1}
\cdot\varepsilon_{\perp2}(n_{-}^\mu n_{-}^\nu+n_{+}^\mu
n_{+}^\nu-n_{-}^\mu n_{+}^\nu-n_{+}^\mu
n_{-}^\nu)\nonumber\\&&+S_{k5}[\varepsilon_{1}\cdot
n_{+}(n_{-}^\nu-n_{+}^\nu)
\varepsilon_{\perp2}^\mu+\varepsilon_{2}\cdot
n_{-}(n_{+}^\nu-n_{-}^\nu) \varepsilon_{\perp1}^\mu]
\nonumber\\&&+S_{k6}(n_{+}^\mu\varepsilon_{1}^\nu
\varepsilon_{2}\cdot n_{-}+n_{-}^\mu\varepsilon_{2}^\nu
\varepsilon_{1}\cdot n_{+})+S_{k7}(n_{-}^\mu\varepsilon_{1}^\nu
\varepsilon_{2}\cdot n_{-}+n_{+}^\mu\varepsilon_{2}^\nu
\varepsilon_{1}\cdot n_{+})\}\ . \eqa
Here,
$T^0=2\frac{N_c^2-1}{16\sqrt{2}N_c^2\sqrt{N_c}m_Q^3}$ and
$g_{\perp}^{\mu\nu}=
 g^{\mu\nu}-\frac{n_{-}^\mu n_{+}^\nu}{2}-\frac{n_{+}^\mu
 n_{-}^\nu}{2}$ comes from the derivative with respect to the
 transverse momentum, such as
 $\frac{\partial k_{\perp}^{\mu}}{\partial
 k_{\perp\nu}}=g_{\perp}^{\mu\nu}$.

The coefficients in above expressions are:
\bqa B_{k1}&&=-\frac{9}{8}\{(2(con^2-1)X_{a}^2
+2\ln(2)(1+con^2+(1-con^2)X_l)-4X_a+\pi^2\} \ , \eqa \bqa
B_{k2}&&=-\frac{9}{32}\{2(X_a^2+(2+4con-2con^2)
X_a-\ln(2)X_l-(1+\ln(2)\nonumber\\&&-4con\ln(2)
+2con^2\ln(2)))+(con^2-2con-1)\pi^2\} \ , \eqa \bqa
B_{k3}&&=\frac{1}{128}\{72\ln(2)
X_l-72X_a^2-144(1-con)^2X_a+(72-288con\nonumber\\&&
+144con^2)\ln(2)-1+36(1-con)^2\pi^2\} \ , \eqa \bqa
B_{k4}&&=\frac{9}{8}\{2(con-1)^2X_a^2
-4(con+2)X_a+2(1-con)\ln(2)X_l\nonumber\\&&+6(1+con)\ln(2)\}\ , \eqa
\bqa B_{k5}&&=\frac{9}{16}(1-con)(2X_a^2-\pi^2)\ , \eqa \bqa
B_{k6}&&=-\frac{9}{16}\{2(conX_a^2-
4(1+con)X_a-2(1+con)(X_l-5)\ln(2))\nonumber\\&&+(3con+4)\pi^2 \}\ ,
\eqa \bqa B_{k7}&&=\frac{9}{16}\{con(-2X_a^2-8X_a+
4(X_l+1)\ln(2)+\pi^2)-8X_a\nonumber\\&&+4(X_l+1)\ln(2)\}\ ,\eqa \bqa
B_{k8}&&=-\frac{9\pi^2}{2}\ ,\eqa \bqa B_{k9}&&=\frac{9\pi^2}{2}\ ,
\eqa \bqa C_{k1}&&=-\frac{9}{16}\{10X_{a}^2-
8X_a+2\ln(2)X_l+5\pi^2+6\ln(2)\}\ , \eqa \bqa
C_{k2}&&=\frac{9}{8}\{6X_{a}^2-2X_a+\pi^2-2\ln(2)\}\ , \eqa \bqa
C_{k3}&&=\frac{9}{64}\{-2X_{a}^2+2X_a-4\ln(2)X_l+3\pi^2-44+2\ln(2)\}\
,\eqa \bqa
C_{k4}&&=-\frac{9}{64}\{10X_{a}^2-6X_a-4\ln(2)X_l-44-2\ln(2)+5\pi^2\}\
,\eqa \bqa C_{k5}&&=\frac{9}{32}\{2X_{a}^2+3\pi^2-8\}\ , \eqa \bqa
C_{k6}&&=-\frac{9}{8}\{2X_{l}\ln(2)-4-\ln(2)\}\ , \eqa \bqa
C_{k7}&&=-\frac{9X_a}{2}-\frac{9\ln(2)X_l}{8}+\frac{45\ln(2)}{8}\ ,
\eqa \bqa C_{k8}&&=\frac{9\pi^2}{8}\ ,\eqa \bqa
C_{k9}&&=-\frac{9}{64}(\pi^2-16)\ , \eqa \bqa
C_{k10}&&=-\frac{9}{64}(\pi^2+16)\ , \eqa \bqa
D_{k1}&&=\frac{9}{8}\{2X_{a}^2-2X_a-\ln(2)X_l+3\ln(2)\}\ , \eqa \bqa
D_{k2}&&=\frac{9}{4}\{X_{l}-1\}\ln(2)\ , \eqa \bqa
D_{k3}&&=-\frac{9}{4}\{X_{a}^2-X_a+\ln(2)\}\ , \eqa \bqa
D_{k4}&&=-\frac{9}{32}\{X_{a}^2+X_a-2\ln(2)X_l-2+\ln(2)\}\ ,
\eqa\bqa
D_{k5}&&=\frac{9}{32}\{7X_{a}^2-5X_a-2X_l\ln(2)-2+7\ln(2)\}\ , \eqa
\bqa S_{k1}&&=\frac{9}{16}\{4\ln(2)X_l-2X_{a}^2-4X_a+\pi^2+4\}\ ,
\eqa \bqa S_{k2}&&=-\frac{9}{8}\{-8X_{a}+\pi^2+2+8\ln(2)\}\ , \eqa
\bqa S_{k3}&&=-\frac{9}{4}\{4X_{a}+(X_l-5)\ln(2)\}\ , \eqa \bqa
S_{k4}&&=\frac{9}{32}\{6X_{a}^2-16X_a+\pi^2+16\ln(2)\}\ , \eqa \bqa
S_{k5}&&=-\frac{9}{32}\{\pi^2-2X_{a}^2\}\ , \eqa \bqa
S_{k6}&&=\frac{9}{32}\{2(X_a^2-4X_a-2\ln(2)X_l+6\ln(2))+3\pi^2\}\ ,
\eqa \bqa S_{k7}&&=-\frac{9}{32}\{4(X_l+1)\ln(2)-X_a^2-8X_a+\pi^2\}\
, \eqa \label{co2}
with $con=1-\frac{f_V (m_1+m_2)}{f_{V}^{T} m_V}.$

For pseudoscalar mesons in final state:
\begin{eqnarray}
{\mathcal{B}}^{\mu\nu} &=& T^{0}\mathbf{B}^{\mu\nu}\ , \label{M-1S0}
\end{eqnarray}
\begin{eqnarray}
\mathbf{B}^{\mu\nu} &=& f_{P}^2\{\frac{\mu_{P}^2}{m_Q^2}
[(M_{k1}g_{\perp}^{\mu\nu})+M_{k2}g^{\mu\nu}+M_{k3}(n_{-}^\mu
n_{-}^\nu+n_{+}^\mu n_{+}^\nu)+M_{k4}(n_{-}^\mu n_{+}^\nu+n_{+}^\mu
n_{-}^\nu)]\nonumber\\&&+M_{k5}g_{\mu\nu}+M_{k6}(n_{-}^\mu
n_{-}^\nu+n_{+}^\mu n_{+}^\nu)+M_{k7}(n_{-}^\mu n_{+}^\nu+n_{+}^\mu
n_{-}^\nu)\}.
\end{eqnarray}
Here $T^0$ is the same as the vector meson case in above, and
\bqa M_{k1}&&=6\{\ln(2)-X_{a}\}\ , \eqa \bqa
M_{k2}&&=2\{X_a-\ln(2)\}\ , \eqa \bqa
M_{k3}&&=\frac{1}{8}\{2X_{a}^2-8X_a+2X_l\ln(2)+1+6\ln(2)\}\ , \eqa
\bqa M_{k4}&&=\frac{1}{4}\{X_a^2+(X_l-1)\ln(2)\}\ , \eqa \bqa
M_{k5}&&=\frac{9\pi^2}{2}\ , \eqa \bqa M_{k6}&&=9-\frac{9\pi^2}{16}\
, \eqa \bqa M_{k7}&&=-\frac{9}{16}\{\pi^2+16\}\ . \eqa \label{co3}

\end{document}